# Assessment of the effectiveness of Omicron transmission mitigation strategies for European universities using an agent-based network model

Jana Lasser[a,b,1], Timotheus Hell[c], and David Garcia[a, b, d]

[a]Graz University of Technology, Institute for Interactive Systems and Data Science, Inffeldgasse 16C, 8010 Graz, Austria; [b]Complexity Science Hub Vienna, Josefstädterstraße 39, 1080 Vienna, Austria; [c]Graz University of Technology, Higher Education and Programme Development, Rechbauerstraße 12, 8010 Graz, Austria; [d]Medical University of Vienna, Center for Medical Statistics, Informatics and Intelligent Systems, Spitalgasse 23, 1090 Vienna, Austria

**Returning universities to full on-campus operations while the COVID-19 pandemic is ongoing has been a controversial discussion in many countries. The risk of large outbreaks in dense course settings is contrasted by the benefits of in-person teaching. Transmission risk depends on a range of parameters, such as vaccination coverage and efficacy, number of contacts and adoption of non-pharmaceutical intervention measures (NPIs). Due to the generalised academic freedom in Europe, many universities are asked to autonomously decide on and implement intervention measures and regulate on-campus operations. In the context of rapidly changing vaccination coverage and parameters of the virus, universities often lack sufficient scientific insight to base these decisions on. To address this problem, we analyse a calibrated, data-driven agent-based simulation of transmission dynamics of 13,284 students and 1,482 faculty members in a medium-sized European university. We use a co-location network reconstructed from student enrollment data and calibrate transmission risk based on outbreak size distributions in education institutions. We focus on actionable interventions that are part of the already existing decision-making process of universities to provide guidance for concrete policy decisions. Here we show that, with the Omicron variant of the SARS-CoV-2 virus, even a reduction to 25% occupancy and universal mask mandates are not enough to prevent large outbreaks given the vaccination coverage of about 80% recently reported for students in Austria. Our results show that controlling the spread of the virus with available vaccines in combination with NPIs is not feasible in the university setting if presence of students and faculty on campus is required.**

COVID-19 | modeling | prevention | network

## Introduction

Many universities face increasing pressure to return to full on-campus operations, while COVID-19 incidence is still high and highly transmissible variants of the virus – such as Omicron – are circulating. A range of simulation studies tried to assess the transmission risk and effectiveness of non-pharmaceutical intervention measures (NPIs) in the university context in the past. These studies have a number of shortcomings that limit their applicability to the decision making of universities. Only a small number of studies consider NPIs in the context of vaccination coverage (1–3) and none considers a situation in which the Omicron variant is dominant. Only a few studies base their models on empirically determined contact networks but in these studies, the transmission dynamics are either not calibrated against empirical data (4–6) or the networks are small (4, 7, 8). Similarly, only a handful of studies calibrate their model parameters against empirically observed outbreaks in educational settings (9–11) but these studies use simulation parameters that were determined for virus variants that are no longer dominant. In addition, existing studies focus on residential colleges and model both contacts in courses and in housing contexts. These studies are applicable to the European higher education sector only to a limited extent, since student housing there differs from the US set-up of in-campus housing and students tend to live spread out in the university's city. As a consequence, COVID-19 prevention policies adopted by European universities have no power to limit social contacts of students outside of university premises. To our knowledge, no existing study combines an empirically determined co-location network with a rigorous calibration of model parameters and simulation scenarios that are relevant for current decision-making processes where the Omicron variant is dominant.

To remedy these shortcomings, we modelled transmission dynamics in a medium-sized European university (TU Graz) with 13,284 students and 1,482 faculty. We base our simulation on an empirically determined co-location network reconstructed from enrolment data from the last term with full on-campus operations in winter 2019/20. At the point of writing, 85% of the students of TU Graz have been vaccinated (12). Based on this high vaccination coverage among students, we investigated whether the university could return to full on-campus operation without risking large outbreaks, even in a situation were the Omicron variant is dominant as is the case as of January 2022.

---

[1]To whom correspondence should be addressed. E-mail: jana.lasser@tugraz.at

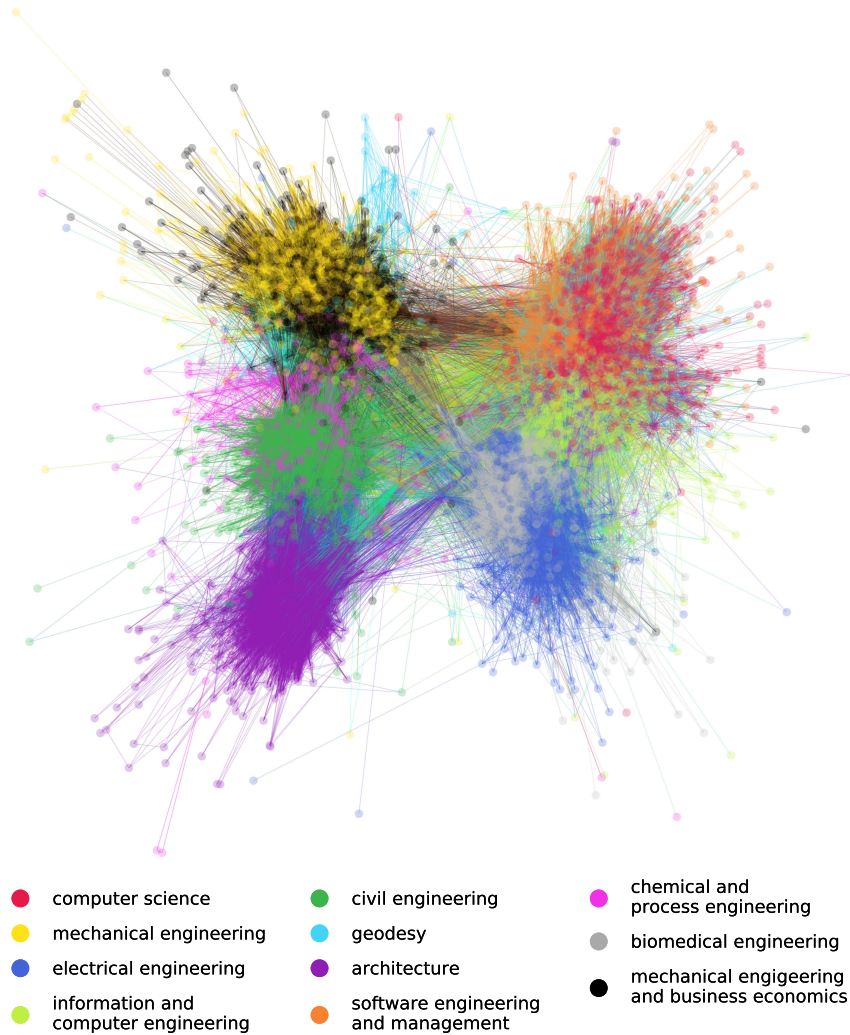

**Fig. 1.** Visualisation of the co-location network of the 11 undergraduate studies of TU Graz. The visualisation shows 5855 students colour-coded by their study and their connections during courses for one week in October 2019.

## 1. Methods

**Co-location network.** We use enrolment data on courses which include lectures, exercises and exams (from now on called "events") from the winter semester 2019/20 to construct the co-location network including students and faculty at TU Graz. The data included a total of 1,752 (mostly semester-long) courses and 1,209 exams attended by students. For every event, a list of dates and locations for the winter semester 2019/20 was available. This resulted in a total of 29,547 unique events at which students and faculty met in the time between October 1 2019 and February 28, 2020. If two students were enrolled in the same event, they were assumed to have contact with each other and the responsible lecturers on every day on which a date for the corresponding event was recorded. If more than one lecturer was responsible for an event, they were also assumed to have contact with each other. The resulting co-location network includes a total of 13,284 students and 1,482 lecturers. See extended methods in the supporting information (SI) for additional details about the co-location network.

**Agent based model.** We simulate the infection dynamics in the university using an agent-based model. The model includes two types of agents: students and lecturers. The model couples in-host viral dynamics with population dynamics. Depending on the viral load over the course of an infection, each agent is in one of five states: susceptible (S), exposed (E), infectious (I), recovered (R) or quarantined (X). In addition, after the presymptomatic phase, agents can stay asymptomatic (I1) or develop symptoms (I2). Agent states and transitions between states are coupled to the in-host viral dynamics and shown in Fig. 2. Agents remain in these states for variable time periods. Every agent has an individual exposure duration (i.e. the time from exposure until an agent becomes infectious), $l$, incubation time (i.e., time until they may show symptoms), $m$, and infection duration, $n$ (i.e., time from exposure until an agent ceases to be infectious). For every agent, we draw values for $l$, $m$, and $n$ from Weibull distributions specified by their mean and standard deviation, ensuring that $1 \leq l \leq m \leq n$. These constraints lead to left-truncated distributions for these epidemiological parameters. The distribution's effective mean and standard deviation are therefore different from the mean and standard deviation specified for the distributions we draw our values from. To address this, we choose the mean and standard deviation of the distributions such that the difference between the effective mean and standard deviation and the mean and standard deviation reported for these epidemiological parameters in the literature are minimal. Specifically, we aim at an effective an incubation time of $3 \pm 1.9$ days (13, 14), exposure duration of $2 \pm 1.9$ days, adapted from the original SARS-CoV-2 strain (15–17), accounting for the shorter incubation time of Omicron. Since to our knowledge no information about the infection duration of Omicron is available yet, we aim for an infection duration of $10.91 \pm 3.95$ days, as reported for the original strain (18, 19). For additional details on the optimisation, see extended methods in the SI. Optimised and effective values for $l$, $m$ and $n$ are reported in Table 1 in the SI.

Infections are introduced into the university through a single, randomly chosen source case that can either be a student or a lecturer. The source case starts in the exposed state on day 0 of the simulation which corresponds to a randomly chosen day of the week. All other agents start in the susceptible state. Recovered agents are assumed to be completely immune to re-infection.

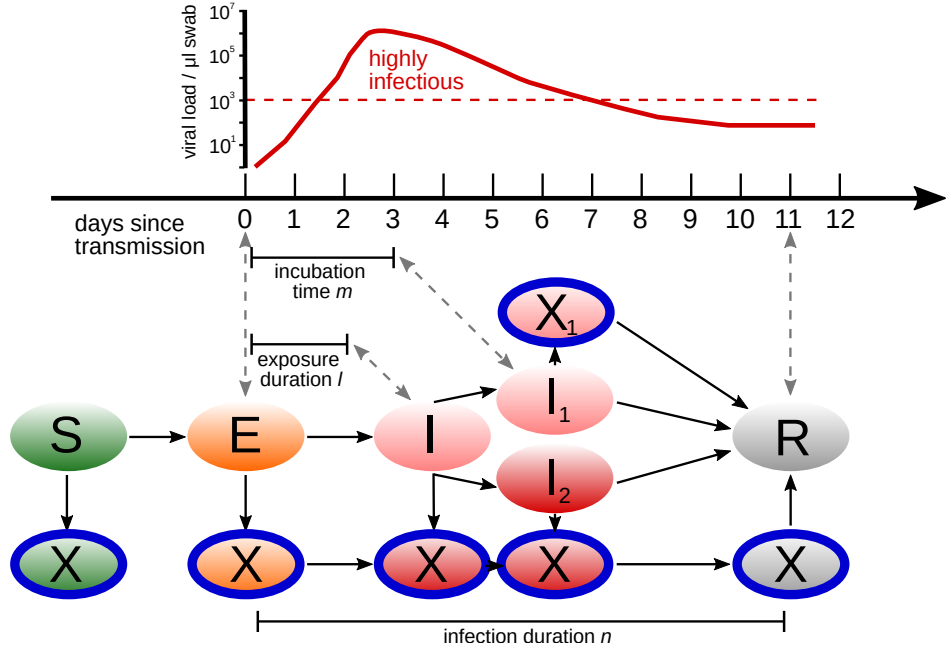

**Fig. 2.** Agents in the epidemiological model can be in the states (ellipses) susceptible (S), exposed (E), infectious (I), infectious without symptoms (I1), infectious with symptoms (I2) and recovered (R). Possible state transitions are shown by black arrows. In each of these states, agents can also be quarantined (X), preventing them from interacting with other agents. Transitions between states follow the development of the viral load in the host sketched above, reproduced from (18).

**Transmissions.** During every interaction, an infected agent can transmit the infection to the agents they are in contact with (specified by the co-location network). Transmission is modelled as a Bernoulli trial with a probability of success, $p$. This probability is modified by several NPIs and biological mechanisms $q_i$, where $i$ labels the measure or mechanism. Here, we consider five such mechanisms: the modification of the transmission risk due to the infection progression ($q_1$), having or not having symptoms ($q_2$), mask wearing of the transmitting and receiving agent ($q_3$ and $q_4$, respectively) and immunisation ($q_5$). Therefore, the probability of a successful transmission is given by the base transmission risk, $\beta$, contact in the university context modified by the combined effect of these five NPIs or biological mechanisms,

$$p = 1 - \left(1 - \beta \prod_{i=1}^{5}(1 - q_i)\right). \quad [1]$$

The modification of transmission risk due to a changing viral load over the course of an infection, $q_1$, is modelled as a trapezoid function that depends on the time an agent has already been exposed to the virus, $t$, given the exposure duration, $l$, incubation time, $m$, and infection duration, $n$, of the infected agent:

$$q_1(t) = \begin{cases} 0 & \text{if } l < t \leq m \\ 1 - \frac{t-m}{n-m+1} & \text{if } t > m \text{ and } t \leq n \\ 1 & \text{else}. \end{cases} \quad [2]$$

If an agent does not show symptoms, transmission risk is reduced by 40% ($q_2$). The arguments for the choices for the values of $q_3$, $q_4$ and $q_5$ are detailed in section "Intervention measures" below.

The base transmission risk $\beta$ is calibrated to reflect the observed transmission dynamics in Austrian secondary schools following Lasser et al. 2021 (20) (see SI section "Calibration" for details). The calibration data was recorded in the same season that our simulation applies to (European autumn and winter), removing the need to explicitly include a seasonal effect. The calibration results in a transmission risk of 2.8% for a university contact for the original virus strain. We adjust this transmission risk to match the Omicron variant, which is about three times as transmissible as the Delta variant (21), which itself is about 2.25 times as transmissible as the original strain (22, 23). This results in a transmission risk of 18.7% per contact for unvaccinated agents in a university setting. This is within the range of reported secondary attack rates (SAR) for omicron in the literature, but most of the literature reports do not explicitly report transmission risks for educational settings and the variance of reported values is high: In a large cohort study from the UK, SAR between unvaccinated non-household contacts was reported to be 10.1% (24). In a large Norwegian cohort study, SAR between household contacts (not differentiated by vaccination status) was reported to be 25.1% (25). In a smaller Spanish contact tracing study, SAR between unvaccinated contacts was reported to be 31.1% in social settings and 36.4% in occupational settings (26). Lastly, a study of a single outbreak in a university setting reports that 38% (10 out of 26 first-level contacts) tested positive (27).

**Intervention measures.** In all simulations, upon first developing symptoms, agents are immediately isolated for 10 days. There are no additional contact tracing and quarantine measures for contacts of the infected agent in place. This reflects the situation currently prevalent in many European countries: vaccinated people or people who were wearing a mask during the contact are only considered low-risk contacts and almost never quarantined. If additional contact tracing measures are introduced, all agents an infected agent had contact with will be quarantined for 10 days two days after the infected agent showed symptoms. This reflects the delay in contact tracing efforts caused by the time it takes for a test result to arrive and to reach all affected contact persons.

For every event location, the seating capacity of the room is known. If occupancy is 100%, all students enrolled in a given event are allowed to attend the event, even if the number of enrolled students surpasses the seating capacity. If occupation is reduced to 50% [25%], students are picked at random and removed from the contact network until 50% [25%] of the seating capacity is reached.

If masks are mandated, all students and lecturers wear masks, which reduces the probability of transmission by 50% if only the infected agent wears a mask ($q_3$) and by 30% if only the receiving agent wears a mask ($q_4$). If both agents wear a mask during the contact, the transmission risk is therefore reduced by 65%. This models the reduction of transmission risk for surgical masks (28) and is in line with a recent review on mask effectiveness (29).

At the beginning of a simulation, 85% of students and lecturers are chosen at random and assigned a "vaccinated" status. This includes immunity from previous infection with other variants of the virus but assumes a population naive to infection with the omicron variant, as was the case in Austria in October 2021. Being vaccinated reduces an agent's chance of getting infected. Since vaccination effectiveness against infection greatly depends on the number of vaccination doses a person has received (30, 31) and wanes with time since the last dose (32), we model different levels of vaccine effectiveness against infection ($q_5$). For the Delta variant, the viral load of vaccinated people was similar to unvaccinated people (33). To our knowledge, similar data does not exist for Omicron yet. We therefore follow the results for the Delta variant and do not assume a lower infectiousness of infected vaccinated agents. Vaccination status and effectiveness does not change throughout the simulation.

## Results

We developed an agent-based model to simulate transmission dynamics on a co-location network determined by the interactions of students and faculty at Graz University of Technology. Figure 1 shows the subset of all contacts from students enrolled in undergraduate programmes at TU Graz, aggregated over the first week of the semester.

We assume a vaccination coverage of 80% among students and staff (12). We study three different lecture hall occupancy levels (100%, 50% and 25%) and two masking mandates (no masks, masks), as well as different vaccine effectiveness against infection (0%, 30%, 50%, 70%). We report distributions of the mean outbreak size (number of infected individuals minus the source case) for 1000 simulations for each scenario. Figure 3 **A** shows the distribution of outbreak sizes for different NPI combinations at a vaccination effectiveness level of 50% – an optimistic estimate of the vaccination effectiveness after two doses of the BNT162b2 or mRNA-1273 vaccines (30). Figure 3 **B** shows the distribution of outbreak sizes for different vaccination effectiveness levels at 25% lecture hall occupancy with masks.

In addition to outbreak sizes, we calculate the average number of secondary infections caused by each agent over the course of their infectious period, $R_\text{eff}$. For our agent-based model, $R_\text{eff}$ is calculated as an individual-based measurement and averaged over all infected individuals in a given simulation, following Breban et al. 2007 (34). Due to the finite size and hegerogeneity of the contact network, $R_\text{eff}$ varies over time. In our simulations most outbreaks do not last longer than 100 days (time steps) and after this time, $R_\text{eff}$ has converged to a stable value (see supplement, Fig. S2). We therefore report $R_\text{eff}$ averaged over the first 100 time steps of the simulation. Distributions of $R_\text{eff}$ for the six different scenarios are shown in Figure 4.

If we assume an optimistic vaccination effectiveness against infection of 50% at a vaccination rate of 85%, with 100% lecture hall occupancy and no masking mandate, the mean outbreak size is 6713 [0; 11181] (95% credible interval) with $R_\text{eff} = 2.0$ [0.0; 3.4]. The maximum observed outbreak size over 1000 simulations is 11270. In 36.0% of the simulation runs the source cases does not infect another person. If both students and lecturers wear masks, the mean outbreak size is reduced to 3342 [0; 8202] and the maximum observed outbreak size is 8335, while $R_\text{eff} = 1.2$ [0.0; 3.0]. On the other hand, if no mask mandate is implemented but instead lecture hall occupancy is reduced to 50%, the mean outbreak size is 4183 [0; 9561] with $R_{\text{eff}} = 1.4\ [0.0; 3.1]$. In a maximum mitigation scenario, if occupancy is reduced to 25% and masks are mandated, mean outbreak sizes are reduced to 4 [0; 29], with $R_{\text{eff}} = 0.4\ [0; 2.9]$, and a maximum observed outbreak size of 517, while 77.3% of the source cases do not lead to another transmission. This shows that even if on average $R_{\text{eff}} < 1$, very large outbreaks are still possible. If 95% instead of 85% of the population is vaccinated, outbreak sizes to not substantially change: on average, 6949 [0; 11148] people are infected with 100% occupancy and no masks, 4290 [0; 9494] with 50% occupancy and no masks, and 5 [0; 52] with 25% occupancy and masks. If vaccine effectiveness is assumed to be 70% – a value that is not realistic with current vaccinations against Omicron – in the maximum mitigation scenario the mean outbreak size drops to 1 [0; 7], with a maximum observed outbreak size of 115 and $R_{\text{eff}} = 0.3\ [0; 2.4]$. With a vaccine effectiveness of 90%, occupancy could be increased to 50% (while still mandating masks), resulting in a mean outbreak size of 1 [0; 13] with $R_{\text{eff}} = 0.3\ [0; 3.0]$ and a maximum observed outbreak size of 55.

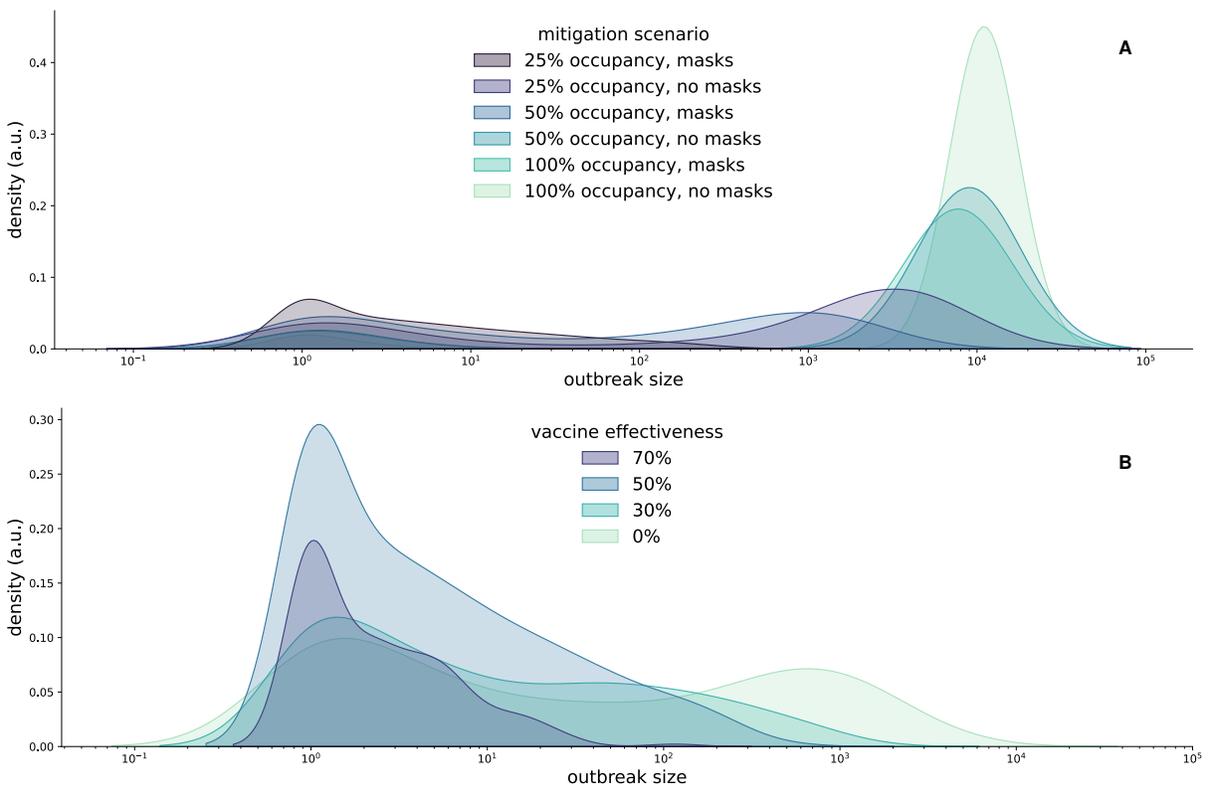

**Fig. 3.** Kernel density estimation of the distribution of outbreak sizes for different mitigation scenarios at a vaccination efficacy against infection of 50% (**A**), and different vaccination effectiveness (**B**) at 25% occupancy with masks. Distributions show outbreak sizes for 1000 simulation runs where the source case infected at least one other agent. Simulation runs where the source case did not cause a secondary infection are excluded.

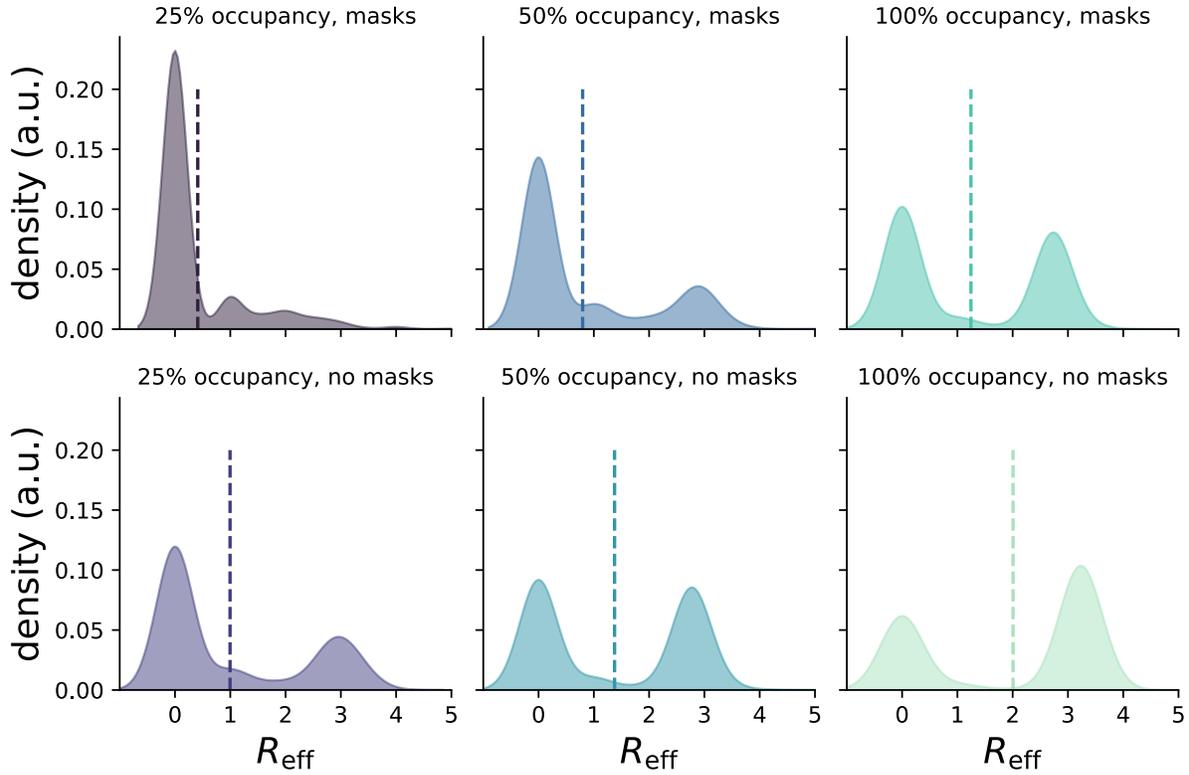

**Fig. 4.** Kernel density estimation of the distribution of $R_{\text{eff}}$ for different mitigation scenarios at a vaccination efficacy against infection of 50%. The dotted lines denote the average of $R_{\text{eff}}$ in each scenario. Distributions show outbreak sizes for 1000 simulation runs where the source case infected at least one other agent.

Our simulations assume testing and isolation of agents as soon as they show symptoms. If additional contact tracing and quarantine measures are introduced, outbreak sizes are slightly reduced, which is consistent with previous findings (8): with 100% occupancy and no masks, on average 6634 [0; 11177] agents are infected. With 50% occupancy and no masks, outbreak sizes are 4182 [0; 9561], and with 25% occupancy and masks 4 [0; 25]. On average, this causes 0.22 quarantine days per student and 0.15 quarantine days per lecturer.

### Discussion

Decreasing lecture hall occupancy to 50% or 25% and imposing masking mandates on students and faculty are two of the most widespread policies to control the spread of COVID-19 adopted by universities (see for example (35)). In Austria, current university policies tie lecture hall occupancy and mask mandates to the level of SARS-CoV-2 community transmission in their region. As Europe enters the cold season, community transmission is increasing or has stabilised on a high level, leading to reduced occupancy and widespread indoor mask mandates in universities. Another NPI that is frequently implemented in institutions and that we investigated in two different studies for schools (20), and nursing homes (36) using the same simulation framework, is frequent preventive testing. With the widespread adoption of so-called "3G" entry rules in Austria, by which students and faculty have to be vaccinated, recently tested, or recovered from a SARS-CoV-2 infection, frequent preventive testing of students has become less desirable, especially given its high cost and implementation complexity. We therefore think that decreasing lecture hall occupancy and mandating masks are the only two feasible options for NPIs that universities can implement.

It is hard to define what an *acceptable* size is for an outbreak, as this depends on many factors, such as the likelihood of causing a severe outcome or long-term damage as well as current hospital and intensive care occupation. The decision of what constitutes too large of a burden for a society is ultimately a political one. We therefore refrain from defining a fixed outbreak size that is acceptable and rather report results for different scenarios.

Our findings suggest that even in a maximum mitigation scenario with only 25% occupancy and a mask mandate, very large outbreaks that reach over 500 students and faculty can happen. The average outbreak size of 4 in this scenario is likely to not be acceptable for university leadership if the goal is to prevent virus transmission within university premises. Given high rates of community transmission, many introductions of the virus in the university setting are to be expected and the occurrence of large outbreaks cannot be ruled out. Given the fact that vaccinations with a higher effectiveness against infection with the Omicron variant are not available yet, this leaves universities with few options: if presence of students and faculty on the university campus is a priority, large outbreaks are likely to be unavoidable. If the prevention of outbreaks is a priority, conducting courses in presence is not a feasible option. If on the other hand vaccines with a higher effectiveness against high viral load become available, our simulations indicate that a high vaccination rate in combination with NPIs can effectively prevent large outbreaks. We note that in an earlier draft version of the present publication, published as a preprint in November 2021 (37) we came to very different conclusions, i.e. that minimal NPI were sufficient to prevent large outbreaks. This was based on the then dominant Delta strain of the virus. The very high transmissibility of Omicron changed the outbreak dynamics completely. This comparison of results show the importance of model calibration and adaptive strategies to the spreading of the virus, where oversimplifications of the spreading dynamics lead to misleading predictions.

Since our model rests on a set of assumptions, an analysis of its shortcomings and the uncertainties associated with these assumptions is warranted. Our model does not include contact situations apart from courses on the university premises. This is a deliberate choice: including such contacts would correspond to multiple concurrent introductions of the infection into the university network, which would further increase the number of transmissions in the university context. Since even without the inclusion of such contacts outbreak sizes are already large, we do not think that adding such contacts would add additional information to our study, while it would introduce a number of additional assumptions to our model. We note that due to this assumption, the outbreak sizes reported here are a lower bound to the outbreak dynamics that are to be expected. We base our contact network on enrolment data of students to courses. Not all students that enrol in a given event show up, as there often is no mandatory attendance, especially under the current circumstances, i.e. an ongoing pandemic. The density of the contact network used in our simulation in the case of 100% occupancy is therefore an upper bound of the number of contacts that are caused by the courses organised by the university.

Since students are on average young (currently 27 years in Austria (38)) and have had to wait a longer time for their first and second vaccine doses in many national vaccination schemes, they will most likely have had their second dose during the summer and most of them will not have had a booster shot yet. We assume the vaccine effectiveness against infection to be 50% for most of our scenarios. This number was reported by Tartof et al. (32) for the BNT162b2 vaccine 4-5 months after the second dose and against the Delta strain of the virus. Vaccine effectiveness against Omicron is still being evaluated at the point of writing, but we assume that 50% effectiveness is very optimistic, as Omicron has demonstrated significant immune escape capabilities (39). If the vaccination effectiveness is assumed to be lower, outbreak sizes increase even further. To account for this uncertainty, we simulated the different mitigation scenarios for different levels of vaccine effectiveness (0%, 30%, 50% and 70%). Therefore, our results should still be applicable once the true effectiveness of currently available vaccines against Omicron are known, or if more effective vaccines become available in the future.

We calibrated our model using empirical observations of cluster sizes in Austrian secondary schools in Autumn 2020 (20). University students tend to have sparser schedules than students in secondary schools, resulting in an overall decreased duration of contact time between students and students, and students and lecturers. On the other hand, the space available per student in university lecture halls is on average smaller than for students in secondary schools (see SI section "Calibration" for details). It is hard to quantify the difference in contact intensity that the difference in contact duration and proximity introduce. Nevertheless, the differences act in different directions and are expected to at least partially cancel each other. We therefore think that it is warranted to assume contact situations in universities and secondary schools are similar enough to use the available data on outbreaks in school settings to calibrate our simulation for an application in the university contact.

Overall, our study assesses the two most common policies to curb the spread of SARS-CoV-2 in European universities: reduction of occupancy and mask mandates, in the context of high vaccine coverage and a dominant high transmissibility variant (Omicron). We find that given the currently reported vaccination rates of students of 85% and above, and an assumed optimistic vaccination effectiveness against infection of 50%, even a maximum mitigation scenario with 25% occupancy combined with a mask mandate for students and lecturers is not enough to prevent large outbreaks.

## Data Archival

The simulation code is published as Python package https://pypi.org/project/scseirx/ (version 1.4.2). The code used to simulate the transmission dynamics at TU Graz is available at . Data used to calibrate the model as well as contact networks and simulation results are available at https://doi.org/10.17605/OSF.IO/UPX7R.

## Notes

The authors declare no competing interests.

DG received funding from the Vienna Science and Technology Fund through the project "Emotional Well-Being in the Digital Society" (Grant No. VRG16-005) for the research leading to these results.

JL received funding from the European Union's Horizon 2020 research and innovation programme under the Marie Skłodowska-Curie grant agreement (No. 101026507 – STAY) for the research leading to these results.

The Open Access publication of this work is supported by TU Graz Open Access Publishing Fund.

**ACKNOWLEDGMENTS.** We thank Susanne Voller from TU IT Services for her help in providing the student enrolment data. We thank Harald Kainz, Rector of TU Graz, for his support of this study.

# Supplementary Information for

## Assessment of the effectiveness of Omicron transmission mitigation strategies for European universities using an agent-based network model

Jana Lasser, Timotheus Hell, David Garcia

jana.lasser@tugraz.at

**This PDF file includes:**

Supplementary text
Figs. S1 to S3
Table S1
SI References

## Supporting Information Text

### Extended methods.

***Co-location network.*** The co-location network used in our simulations includes a total of 13,284 students and 1,482 lecturers. According to official university statistics (1), TU Graz had 15,909 enrolled students at the end of the winter semester 2019/20. The difference to the students included in our co-location network can be explained by the large number of students who are enrolled but not active – a common phenomenon at Austrian universities[*]. Regarding its number of students, TU Graz is a medium-sized university, comparable to other Austrian universities: on average, Austrian universities have 13,797 enrolled students[†] (1). For other countries in Europe, larger countries tend to have more students per university. For example the average number of students in German universities was 16,628 in the winter semester 2019/20 (2), in Italy it was 19,458 in 2012 (3) and in Switzerland it was 10,545 (numbers per university from the years 2017 to 2021) (4). We conclude that with 15,909 enrolled students, TU Graz is comparable in size to the average European university.

On weekdays, students [lecturers] have an average node degree (i.e. number of contacts) of 19.2 [23.2]. Degree distributions for students and lecturers are shown in Fig. S1. As qualitatively shown in Fig. 1 of the main text, the network includes a number of communities represented by the individual degree programmes. This is also reflected in the modularity score of 0.36, calculated with the "greedy modularity" algorithm following Clauset et al. 2004 (5). We use the implementation of the algorithm provided by the Python networkx package version 2.7.1[‡].

We note that a substantial number of students (3257) contained in the network are enrolled in a "Jointly Offered Study Programme" ("NAWI Graz") that features lectures both at TU Graz and University of Graz. These students visit only about 50% of the lectures that are part of their degree at TU Graz and our data does not include information about the lectures provided by University of Graz. This is also reflected in their average node degree, which is only 8.3. The transmission dynamics are substantially dampened if only NAWI Graz students are considered. Overall we note that including NAWI Graz students in the simulations yields very similar results as excluding them.

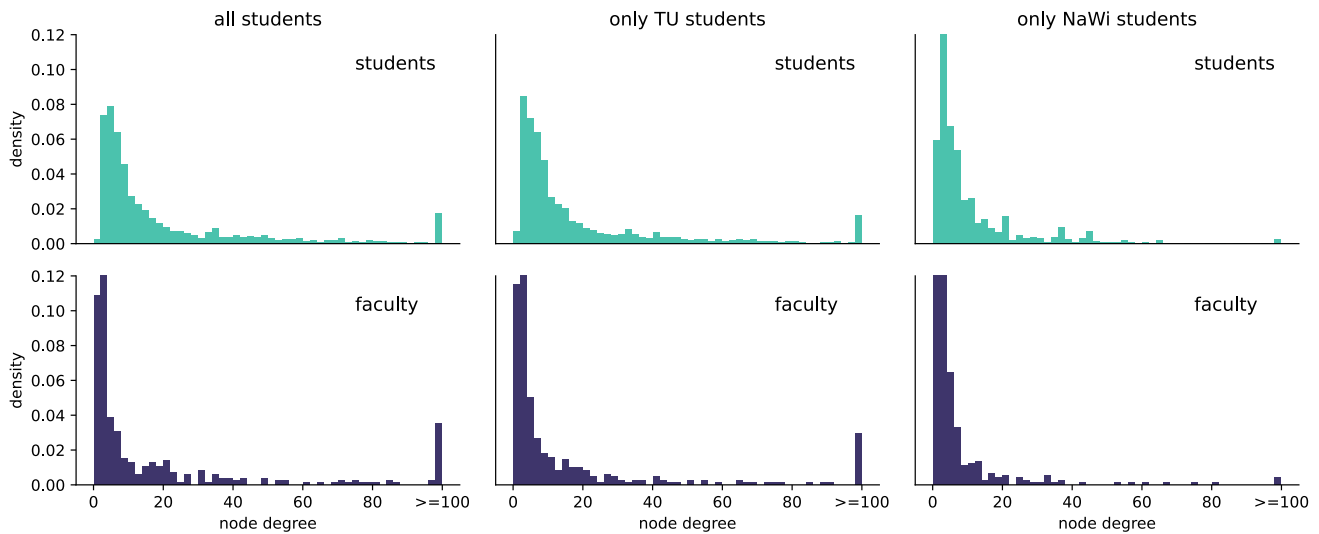

**Fig. S1.** Degree distributions for students (top row) and faculty (bottom row) if all students are included in the network (left), only students that are enrolled in a TU degree programme are included (middle) or only students that are enrolled in a joint degree with Uni Graz (NAWI) are included (right).

***Optimisation of epidemiological parameters.*** To ensure the logical consistency of the combination of the exposure duration $l$, incubation time $m$ and infection duration $n$ within every individual agent, we impose the constraints $1 \leq l \leq m \leq n$. By imposing these constraints, we left-truncate the corresponding distributions for these epidemiological parameters. Therefore, the effective mean and standard deviation of the distributions is different from the mean and standard deviation that defines the distributions we draw our values from. To nevertheless match the values for these epidemiological parameters reported in the literature, we optimise the mean and standard deviation used to define the Weibull distributions we draw our parameters from such that the effective mean and standard deviation after left-truncating the distributions matches the values reported in the literature as closely as possible. To achieve this, we first perform a coarse grid search over the six-dimensional parameter space (start : stop : step) spanned by the means $\mu_l \in [1.4 : 4.0 : 0.2]$, $\mu_m \in [1.0 : 3.0 : 0.2]$ and $\mu_n \in [10.0 : 11.4 : 0.2]$, and standard deviations $\sigma_l \in [2.0 : 3.0 : 0.25]$, $\sigma_m \in [1.5 : 2.5 : 0.25]$ and $\sigma_n \in [3.0 : 5.0 : 0.5]$ of the three distributions. For each of the 154000 possible parameter combinations we drew values for $m$, $n$ and $l$ from the corresponding distributions 1,000 times, computed the effective mean and standard deviation of the distributions and calculated the difference to the literature values. The squared sum of difference was then used to determine the distance of a given parameter combination from the literature values.

---

[*] According to the Austrian ministry of Education and research, up to 42% of students enrolled in Austrian universities are not actively pursuing their studies. See https://www.bmbwf.gv.at/Themen/HS-Uni/Hochschulgovernance/Leitthemen/Qualit%C3%A4t-in-der-Lehre/Pr%C3%BCfungsaktivit%C3%A4t.html, accessed 2022-01-16

[†] the average excludes the University for Continuing Education in Krems, which only has 19 students

[‡] See https://networkx.org/documentation/stable/reference/algorithms/generated/networkx.algorithms.community.quality.modularity.html

After the coarse grid search, we performed a second grid search with a finer-grained step-size around the value with the smallest difference to the literature values found in the coarse grid search. Again, we searched the full parameter space spanned by $\mu_l \in [3.5 : 4.4 : 0.05]$, $\mu_m \in [1.55 : 2.05 : 0.05]$ and $\mu_n \in [10.65 : 10.85 : 0.1]$, and standard deviations $\sigma_l \in [2.75 : 3.35 : 0.1]$, $\sigma_m \in [1.75 : 2.35 : 0.1]$ and $\sigma_n \in [3.75 : 4.25 : 0.1]$ for a total of 184338 possible parameter combinations. This time, we drew values for $l$, $m$ and $n$ 10,000 times from the corresponding distributions.

We concluded the optimisation process with a third grid search with an even finer-grained step-size around the value with the smallest difference to the literature values found in the fine grid search. We searched the full parameter space spanned by $\mu_l \in [4.00 : 4.31 : 0.01]$, $\mu_m \in [1.40 : 1.70 : 0.01]$ and $\mu_n = 10.65$, and standard deviations $\sigma_l \in [2.75 : 3.05 : 0.1]$, $\sigma_m \in [1.75 : 2.05 : 0.1]$ and $\sigma_n \in [4.15 : 4.35 : 0.1]$ for a total of 47616 possible parameter combinations. This time, we drew values for $l$, $m$ and $n$ 50,000 times from the corresponding distributions.

We note that we followed a similar approach to determine the optimal values for the epidemiological parameters of the wild type strain used in the calibration of the model. The most optimal values, effective values after left-truncating and target literature values for the means and standard deviations of the epidemiological parameters of the wild type and omicron variant are reported in Table S1. The left-truncated distributions for $l$, $m$ and $n$ for the omicron variant are shown in Fig. S2.

| variable | optimal value | effective value | literature value |
|---|---|---|---|
| wild type | | | |
| $\mu_l$ | 6.8 | 4.85 | 5.0 |
| $\mu_m$ | 6.3 | 6.48 | 6.4 |
| $\mu_n$ | 8.65 | 10.99 | 10.91 |
| $\sigma_l$ | 2.75 | 1.56 | 1.9 |
| $\sigma_m$ | 0.95 | 0.91 | 0.8 |
| $\sigma_n$ | 4.75 | 3.87 | 3.95 |
| omicron | | | |
| $\mu_l$ | 4.31 | 2.01 | 2.0 |
| $\mu_m$ | 1.49 | 3.16 | 3.0 |
| $\mu_n$ | 10.65 | 10.94 | 10.91 |
| $\sigma_l$ | 3.05 | 1.30 | 1.9 |
| $\sigma_m$ | 1.75 | 2.06 | 1.9 |
| $\sigma_n$ | 4.15 | 4.06 | 3.95 |

**Table S1.** Values chosen for the means $\mu$ and standard deviations $\sigma$ defining the Weibull distributions from which we draw the epidemiological parameters exposure duration $l$, incubation duration $m$ and infection duration $n$ for the agents in our simulation. For each parameter, the optimal value found in the grid search, the effective value after left-truncating the distributions and the target literature value are given.

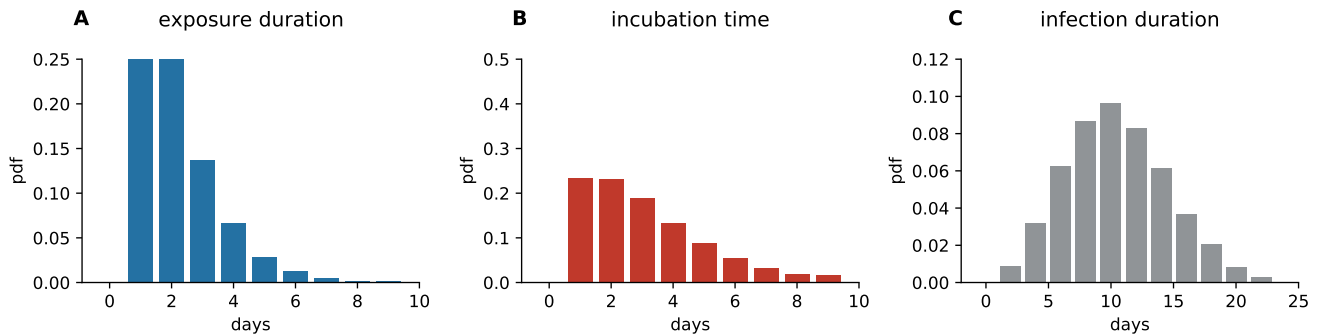

**Fig. S2.** Left-truncated distributions of parameters used in the agent based simulation for the omicron variant. A exposure duration, $l$, B incubation time, $m$, and C infection duration, $n$.

***Calibration.*** Since there is no data of sufficiently high quality about outbreaks in a higher education setting available to us, we resort to calibrating our transmission dynamics against outbreaks observed in Austrian secondary (children aged 12 to 18) and upper secondary schools (children aged 14 to 18). We follow a similar calibration approach to the version that was already described in Lasser et al. 2021 (6), adapted to only include secondary and upper secondary schools. We note that we follow a similar approach as described in section "Optimisation of epidemiological parameters" to recover the effective distributions after left-truncation for the wild type variant. Epidemiological parameters used for the wild type variant in the calibration simulations are reported in Table S1.

To perform the calibration, we first calibrate the transmission risk $\alpha$ between two adults living in the same household to match the household secondary attack rate that was reported to be 28.3% (7) for the wild type variant. This leads to a transmission risk of $\alpha = 5.9\%$ per contact between two adults living in the same household where one adult is infected.

Household contacts are not the same as contacts in an educational setting. We therefore need to determine the reduction of transmission risk $\gamma$ of a contact in an educational setting relative to a household contact. The transmission risk $\beta$ used in our simulations is then computed as the product between the household transmission risk and the transmission risk reduction in an educational setting $\beta = \alpha\gamma$. To this end we perform simulations of outbreaks in schools, relying on our modelling of the transmission dynamics in schools that has been published elsewhere (6).

To calibrate our model, we perform simulations that match the situation in Austrian schools at the time the calibration data was collected. The data used to calibrate our model was collected in calendar weeks 35-46 2020 in Austrian schools. The data includes 116 distinct outbreaks with a total of 270 cases in upper secondary schools, and 70 distinct outbreaks with a total of 580 cases in secondary schools. At the time of data collection there was no mask mandate in Austrian schools and mask usage was rare. We therefore do not include any mask adherence parameter or non-mask periods in our calibration simulations. In addition, activities in Austrian schools were very limited during that time: music and sports courses as well as trips were suspended, and beginning and end times of courses as well as breaks were staggered, such that different cohorts would not come into contact with each other.

We compare the empirically observed and simulated distribution of cluster sizes in schools to find the best value for $\gamma$. Specifically, we optimise the sum of the squared differences between the empirically observed cluster size distribution and the cluster size distribution from simulations $e_1$, and the sum of squared differences between the empirically observed ratio of infected students versus infected teachers, and the simulated ratio $e_2$. We perform simulations for values of $\gamma$ in the range of 0.32 and 0.63, with steps of 0.01, conducting 10000 simulations per parameter value and school type (secondary, upper secondary). We calculate the overall difference between the simulated and empirically observed cluster characteristics as

$$E = \sum_i \frac{e_{1,i} + e_{2,i}}{N_i} \, ,$$

where $N_i$ is the number of empirically observed clusters for school type $i$. We have shown in previous work (6) that the outcome of the calibration is robust to the choice of distance metric.

We then fit a second-order polynomial to the resulting error terms $E$ for different choices of $\gamma$. We find that a value of $\gamma = 0.47$ minimises $E$. This means that contacts in the school context have a transmission risk of $5.9\% \cdot 0.47 = 2.8\%$ for the original strain of the virus. Since Omicron is estimated to be about 6.75 times as transmissible as the original strain, the final base transmission risk in our simulation is set to be $\beta = 18.5\%$.

In our school study (6), we find that age reduces transmission risk by a negligible amount in school children (0.005 per year younger). To our knowledge, there is no evidence that transmission risk significantly depends on age for adults. We therefore think it is warranted to omit the age dependence of transmission risk in the university setting. University contacts are more likely to be shorter than school contacts, since the majority of lectures, tutorials or exams last for 1-3 hours and the mean time a student at TU Graz spends in lectures and exams every day is 165 minutes. On the other hand, students in Austrian upper secondary and secondary schools spend on average 5–6 hours with each other in the same classroom. The average space available for a student in a classroom of an Austrian upper secondary [secondary] school is 2.2 [2.1] m$^2$, respectively, based on the average number of 23 [24] students per class (8) and the minimum room size of 50 m$^2$, mandated by Austrian school building regulations (9). The average space available for a student in a lecture hall at a university (if all available seats are occupied) is 1.9 m2. The reduced average contact time is expected to reduce transmission risk, whereas the reduced distance between students is expected to increase transmission risk. Based on these facts, we assume that contacts between students, and students and students and teachers in the context of Austrian upper secondary and secondary schools are sufficiently similar to contacts between students, and students and faculty in Austrian universities.

**Calculating the reproduction number, $R_{\text{eff}}$.** We use the reproduction number calculated from infection chains in our model to report and compare outcomes of different intervention scenarios. Since our model is agent-based, $R_{\text{eff}}$ is calculated as an individual-level measure (10) by counting the number of secondary infections a focal individual has caused. Therefore, in the context of our agent-based model, $R_{\text{eff}}$ is not a model control parameter but rather an observable. In contrast to the commonly discussed reproductive number $R_0$, $R_{\text{eff}}$ can be < 1 on average, while still observing large outbreaks. Since the number of agents in our model is limited, we expect the finite size of the model to influence the number of secondary infections caused by each agent as the infection spreads through the system, since the pool of susceptible agents is not infinite and depletes with time. Specifically, we expect $R_{\text{eff}}$ to depend on the number of simulation time-steps $t$ included in it's calculation. Figure S3 shows the values of $R_{\text{eff}}$ calculated at different time-steps $t$. As shown in the figure, $R_{\text{eff}}$ first increases as the infection spreads into dense areas of the network and then drops again, as the number of susceptible agents is depleted. Around $t = 100$, $R_{\text{eff}}$ has converged to a stable value. This pattern emerges for all mitigation scenarios and all values chosen for the vaccination effectiveness. We use the value for $R_{\text{eff}}$ calculated after 100 time steps to compare different simulated scenarios.

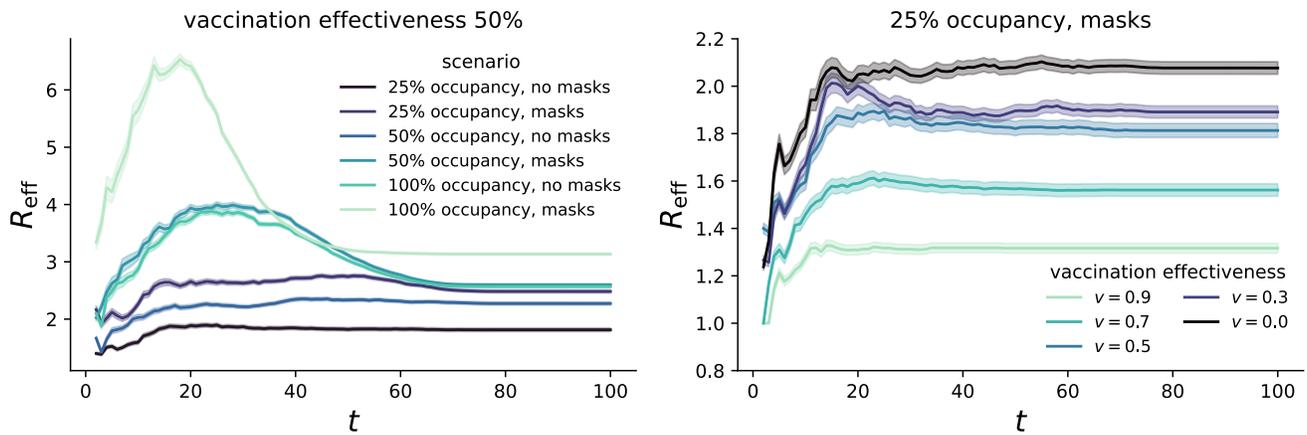

**Fig. S3.** Average value of $R_{\text{eff}}$ and it's standard error (shaded area) for different scenarios and different vaccination effectiveness. The value of $R_{\text{eff}}$ depends on the number of simulation time-steps $t$ included in the average.